\documentclass{article}
\usepackage{graphicx}
\usepackage{setspace}
\usepackage[margin=1.3in]{geometry}
\usepackage[backend=biber, style=apa, natbib=true]{biblatex}
\bibliography{references}
\usepackage{hyperref}
\usepackage{authblk}
\usepackage{tabularx}
\usepackage{footnotebackref}
\usepackage{float}
\usepackage[labelfont=bf]{caption} 
\newcommand{\figref}[1]{\hyperref[fig:#1]{\textbf{Fig.~\ref*{fig:#1}}}}
\newcommand{\tableref}[1]{\hyperref[tab:#1]{\textbf{Table~\ref*{tab:#1}}}}

\newcommand{\insertfig}[3]{
\begin{figure}[h]
\centering
\begin{minipage}{0.75\textwidth}
    #1
    \caption{#2}
    \centering
    \label{#3}
\end{minipage}
\end{figure}
}

\title{Exploring a Cognitive Architecture for\break Learning Arithmetic Equations\vspace*{10pt}}
\author{Cole Gawin}
\affil{University of Southern California}
\affil{\texttt{gawin@usc.edu}}
\date{May 2024}

\begin{document}
\onehalfspacing
\setlength{\parskip}{6pt}

\widowpenalty=10000
\clubpenalty=1000000

\maketitle

\begin{abstract}
The acquisition and performance of arithmetic skills and basic operations such as addition, subtraction, multiplication, and division are essential for daily functioning, and reflect complex cognitive processes. This paper explores the cognitive mechanisms powering arithmetic learning, presenting a neurobiologically plausible cognitive architecture that simulates the acquisition of these skills. I implement a number vectorization embedding network and an associative memory model to investigate how an intelligent system can learn and recall arithmetic equations in a manner analogous to the human brain. I perform experiments that provide insights into the generalization capabilities of connectionist models, neurological causes of dyscalculia, and the influence of network architecture on cognitive performance. Through this interdisciplinary investigation, I aim to contribute to ongoing research into the neural correlates of mathematical cognition in intelligent systems.
\end{abstract}

\section{Introduction}
Understanding how humans learn to perform arithmetic is a compelling aspect of cognitive science with far-reaching implications for education, cognitive psychology, and artificial intelligence. Across cultures and ages, the ability to perform basic arithmetic computations like addition or division is essential for everyday functioning. The neurological mechanisms underlying arithmetic ability learning remain a topic of intense investigation, particularly in the context of computational models that seek to simulate and elucidate these processes.

One lens through which researchers analyze cognitive processes is through the development of cognitive architectures. As described by the USC \citet{usc_cognitive_architecture}, cognitive architectures are computational frameworks that attempt to capture the structure and function of the human mind, providing a theoretical basis for understanding cognitive processes such as perception, memory, language, and reasoning. These architectures serve as blueprints for building computational models that simulate human-like cognitive abilities, offering insights into the underlying mechanisms of cognition.

Within the realm of cognitive architectures, connectionist models\footnote{Connectionist models are often referred to as \textit{artificial neural networks} (ANNs) in the field of computer science. This term marks a comparison to \textit{biological} neural networks found in the human brain.} stand out for their ability to capture the emergent properties of complex systems through the interaction of simple processing units, analogous to neurons in the brain. By representing arithmetic knowledge as patterns of activity distributed across interconnected nodes, these models can simulate the long-term memory storage of arithmetic operations, shedding light on the underlying mechanisms that support this critical ability.

In this paper, I produce a neurobiologically plausible cognitive architecture for learning arithmetic operations. Specifically, I explore a hypothetical framework through which an intelligent system can encode numerical information as sparse distributed representations and memorize addition tables. I do so by implementing a novel number vectorization embedding network, as well as a connectionist associative memory model. In training such a model on arithmetic operations, I aim to elucidate the mechanisms underlying mathematical learning and cognition.

By systematically investigating topics such as generalization, neural substrates of dyscalculia, and the impact of network architecture, I indend to deepen the field's appreciation of how individuals acquire, represent, and utilize arithmetic knowledge. This interdisciplinary approach integrates insights from cognitive psychology, neuroscience, and artificial intelligence. I endeavor to bridge the gap between cognitive theory and computational modeling, thereby advancing both our theoretical understanding of arithmetic learning and the practical applications of connectionist models in education, psychology, and AI.

\section{Background \& Literature Review}

\subsection{Knowledge Acquisition via Rote Memorization}

Rote memorization, often characterized as the repetitive learning of information without understanding its underlying meaning, has long been a traditional method for acquiring knowledge in various domains, including mathematics. Though this approach is of intense debate regarding its neglect towards fostering deep understanding and critical thinking skills, it remains prevalent in educational settings due to its practicality in aiding recall and mastery of factual information \parencite{Brown2014-yz}.

In the context of mathematics, rote memorization plays a pivotal role in establishing a foundation of mathematical thinking. For instance, ``skill-and-drill'' instruction has long been a widely accepted methodology for the acquisition of basic arithmetic knowledge. Specifically, teachers often employ tools like addition and multiplication tables to have students memorize answers to basic equations. Since argued as the optimal technique to instill this knowledge into students' long-term memory by \citet{Thorndike1921}, this has been the primary method used by teachers in early education to teach arithmetic. 

\subsection{Supervised Learning}

Supervised learning is a fundamental paradigm in machine learning where a model learns to map input data to corresponding output labels based on example pairs provided during training. This approach is akin to a teacher providing labeled examples to a student, allowing them to learn to associate input patterns with correct outputs.

In supervised learning, the goal is to approximate or learn the underlying mapping function $f$ that relates input features $X$ to output labels $y$. Mathematically, this relationship can be represented as $y = f(X) + \epsilon$, where $\epsilon$ denotes random error. The model aims to minimize the discrepancy between its predictions and the true labels in the training data. This is done by training models through numerous epochs of training to learn from the mistakes it made on each trial.

The interactive activation and competition (IAC) model developed by Rumelhart and McClelland \parencite*{rumelhart1986} provides a cognitive perspective on supervised learning. Inspired by the architecture of the human brain, the model employs feedback mechanisms to drive learning and refine representations overtime. This process of error-driven learning mirrors the principles of supervised learning, where feedback from correct and incorrect predictions guides the adjustment of model parameters to minimize errors.

\subsection{Connectionist Models and Associative Memory}
Connectionist modeling has garnered significant attention in cognitive science for their ability to simulate complex cognitive processes. These models are inspired by the structure and function of the brain, consisting of interconnected nodes (neurons) that process and transmit information through weighted connections. One fundamental aspect of connectionist models is their capacity to encode and retrieve information through associative memory mechanisms.

Inspired by the distributed and parallel processing observed in biological neural networks, connectionist models offer a biologically plausible framework for studying cognitive phenomena \parencite{rumelhart1986}. Owing to their ability to learn from examples via trial and error and identify non-trivial patterns in information, these models can be particularly well-suited for recreating neurological tasks involving pattern recognition, learning, and memory.

A common implementation of connectionist models is associative memory networks. Associative memory involves forming connections between patterns or concepts, thereby enabling the retrieval of information based on learned associations. This property mirrors the functioning of human memory in the hippocampus, specifically the parietal-hippocampal network \parencite{wagner2005}, where activation of one concept can trigger the retrieval of related information.  

\subsection{Sparse Distributed Representations}

Sparse distributed representations (SDRs) have also been a topic of interest in neural network models. In SDRs, information is represented by a binary vector where only a small fraction of the elements are active (typically around 15\%), while the majority remain inactive. Despite their sparsity, SDRs can represent a large number of unique patterns due to the potential for overlap between different active bits. This is in contrast to one-hot encodings, in which each individual element of a set is represented as a single active element in a vector; one-hot encodings lack the efficiency in representations present in SDRs and suffer from high dimensionality, making them improbable in biological neural networks.

\citet{Barlow1972} suggested sparseness as a critical aspect in neural representations of stimuli—specifically, that neural systems tend to represent information using the smallest possible number of active neurons. This is present for sensory stimuli as well as encoding throughout the cerebral cortex, especially in the hippocampus. More recent research \parencite{Ahmad2015PropertiesOS} investigated the importance of SDRs in cortical encoding, and found that SDRs are crucial to higher order memory. This could explain how the hippocampus, a structure in the brain associated with memory and learning, is able to encode semantic memories through forming distributed representations that capture the relationships between concepts \parencite{McClelland1995}. Save for extreme theories such as the ``grandmother cell'' hypothesis \parencite{Barwich2019}, it is generally accepted that sparse distributed representations offer the most efficient and effective means of encoding information in both biological neural networks and deep learning models. 

A prominent application of SDRs in the realm of machine learning is word vectorization techniques such as \texttt{word2vec}, introduced by \citet{Mikolov2013EfficientEO}. Word embeddings have revolutionized natural language processing (NLP) tasks by encoding semantic and syntactic information into vector representations. Traditional approaches to representing words in NLP, like one-hot encodings, suffer from high dimensionality and lack of semantic relationships between words. In contrast, word embeddings capture semantic similarities between words by placing them in a continuous vector space where similar words are represented by vectors that are close in proximity. This is akin to how the hippocampus and cortex form and store connections between related memories.

\subsection{Learning Mechanisms}

The most commonly employed learning mechanism in ANNs is backpropagation. Backpropagation involves calculating gradients of the loss function with respect to the network's weights and then adjusting those weights accordingly to minimize the error \parencite{Rumelhart1986-lr}.

However, backpropagation is generally considered to be biologically implausible.\footnote{Hinton has himself recognized this shortcoming \parencite*{hinton2007}, and fellow AI forefather Yoshua Bengio and colleagues have investigated from a computer science perspective how deep learning models could be developed using biologically plausible mechanisms \parencite*{Bengio2015TowardsBP}.} This process requires knowledge of the network's output and an explicit error signal, which is not readily available in the brain. Moreover, the precise and synchronized adjustments of weights across layers, as backpropagation entails, are not biologically realistic. The brain operates in a more distributed and parallel manner, with learning mechanisms that are inherently local and activity-dependent \parencite{OReilly1996-ub}.

One alternative framework that strives for greater biological plausibility is the Leabra algorithm (Local, Error-driven and Associative, Biologically Realistic Algorithm), proposed by \citet{oreillyphd}. Leabra aims to capture key principles of neural computation observed in the brain \parencite{OReilly2015-zt}. Unlike backpropagation, Leabra emphasizes local learning rules, where each neuron's synaptic weights are updated based on local information and without the need for explicit error signals.

In Leabra, learning occurs through a combination of Hebbian learning, which strengthens connections between neurons that are simultaneously active, and error-driven learning, which adjusts synaptic weights based on the difference between actual and expected outcomes. This hybrid approach mimics the interplay between associative learning and error correction mechanisms observed in biological systems.

\section{Methodology}

\subsection{Cognitive Architecture Design}

For this study, I developed a cognitive architecture that consists of two simuations:

\begin{enumerate}
    \item A \textbf{numerical embedding network} that learns to develop sparse distributed representations of numbers 0–81.
    \item A \textbf{associative memory model} implemented with a connectionist model that learns to associate equation inputs with corresponding outputs.
\end{enumerate}

Both of these components necessitated the development of separate simulation models. The associative memory model is trained using inputs developed by the numerical embedding network, hence these simulations were trained sequentially.

These simulations were developed using the Emergent software system \parencite{Aisa2008-zj}, which implements the Leabra learning algorithm \parencite{oreillyphd}. These systems allow for the development of biologically plausible neural networks, a critical aspect of computational models of cognitive processes.

\subsection{Model Implementations}

\subsubsection{Numerical Embedding Network}

\insertfig{\includegraphics[width=\textwidth]{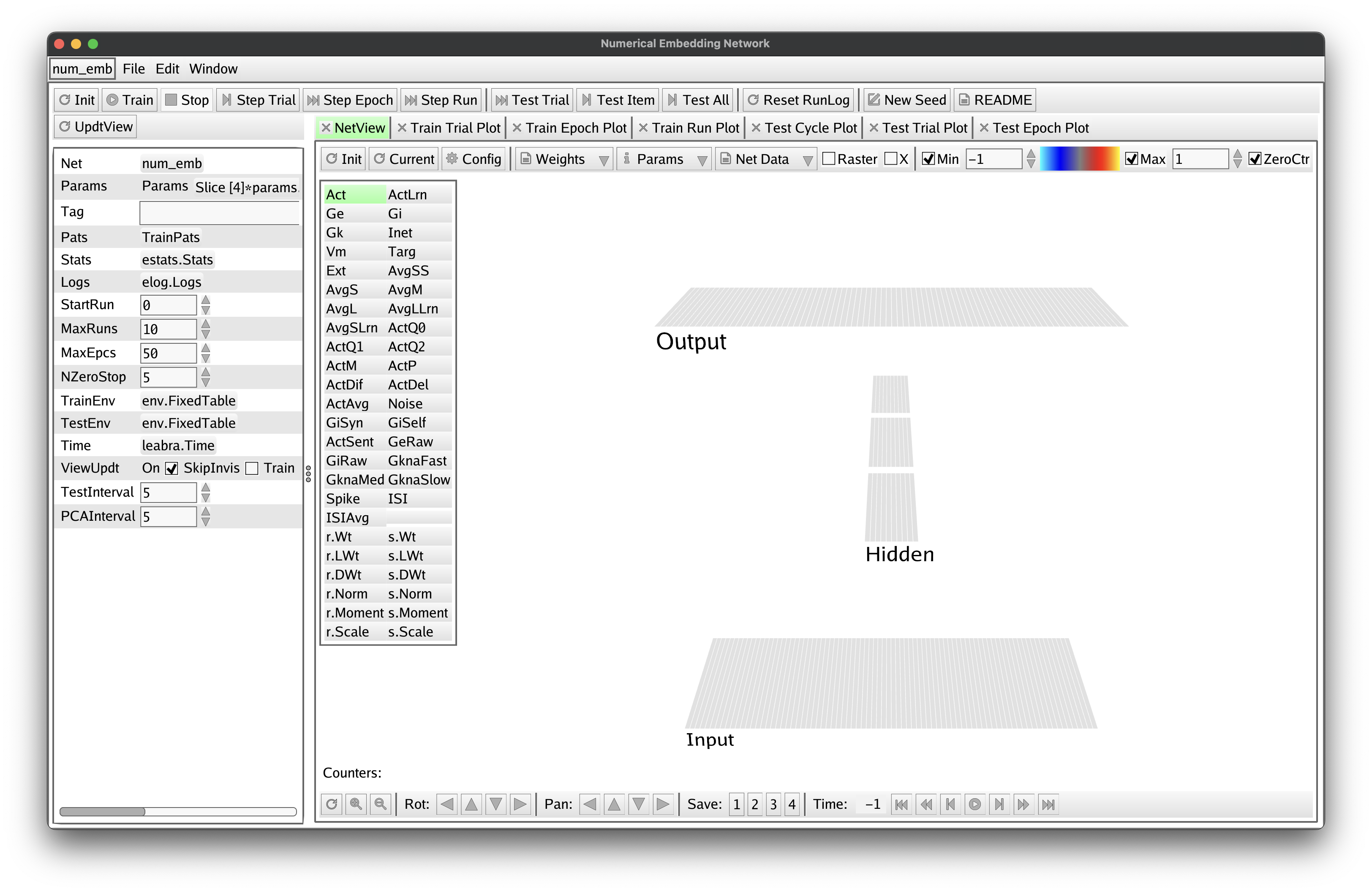}}{Numerical embedding network simulation.}{fig:1}

The numerical embedding network has three layers: two one-hot encoding layers that represent the number being encoded, and one layer that generates the sparse distributed representations of the number. The one-hot encoding layers are size $1\times82$ since 82 numbers (0-81) are being encoded, and the SDR layer is size $3\times10$ representing 30 units total. Both of the first two layers are bidirectionally connected to the SDR layer. This results in a network in which, when one one-hot encoding layer is provided with a specific input, the activations in the SDR layer that result from its learned weights will activate the same input in the second one-hot encoding layer.

The activations in the SDR layer for each of the numbers the network is trained on are then considered the number's \textit{embeddings}. The SDR layer being smaller than the input layers means this process is a method of dimensionality reduction. Moreover, this property allows for the weights it learns to result in sparse activations. This approach to generating embeddings in a lower-level vector space than the inputs was popularized by cognitive psychologist and AI pioneer Geoffery Hinton and computer scientist Russ Salakhutdinov in their seminal paper, \textit{Reducing the Dimensionality of Data with Neural Networks} \parencite*{Hinton2006}.\footnotemark

\footnotetext{Later work by \citet{sutskever2014sequence} in \textit{Sequence to Sequence Learning with Neural Networks} extended this concept to the domain of sequence-to-sequence learning, which is employed within encoder/decoder generative language models.}

These embeddings are used as inputs and outputs in the associative memory network, allowing a relatively small neural network to be trained on data using a large set of possible numbers. Furthermore, the network (theoretically) wouldn't have to reconfigure its size when more inputs/outputs are incorporated since the size of the embeddings is a fixed value.

\subsubsection{Associative Memory Network}

\insertfig{\includegraphics[width=\textwidth]{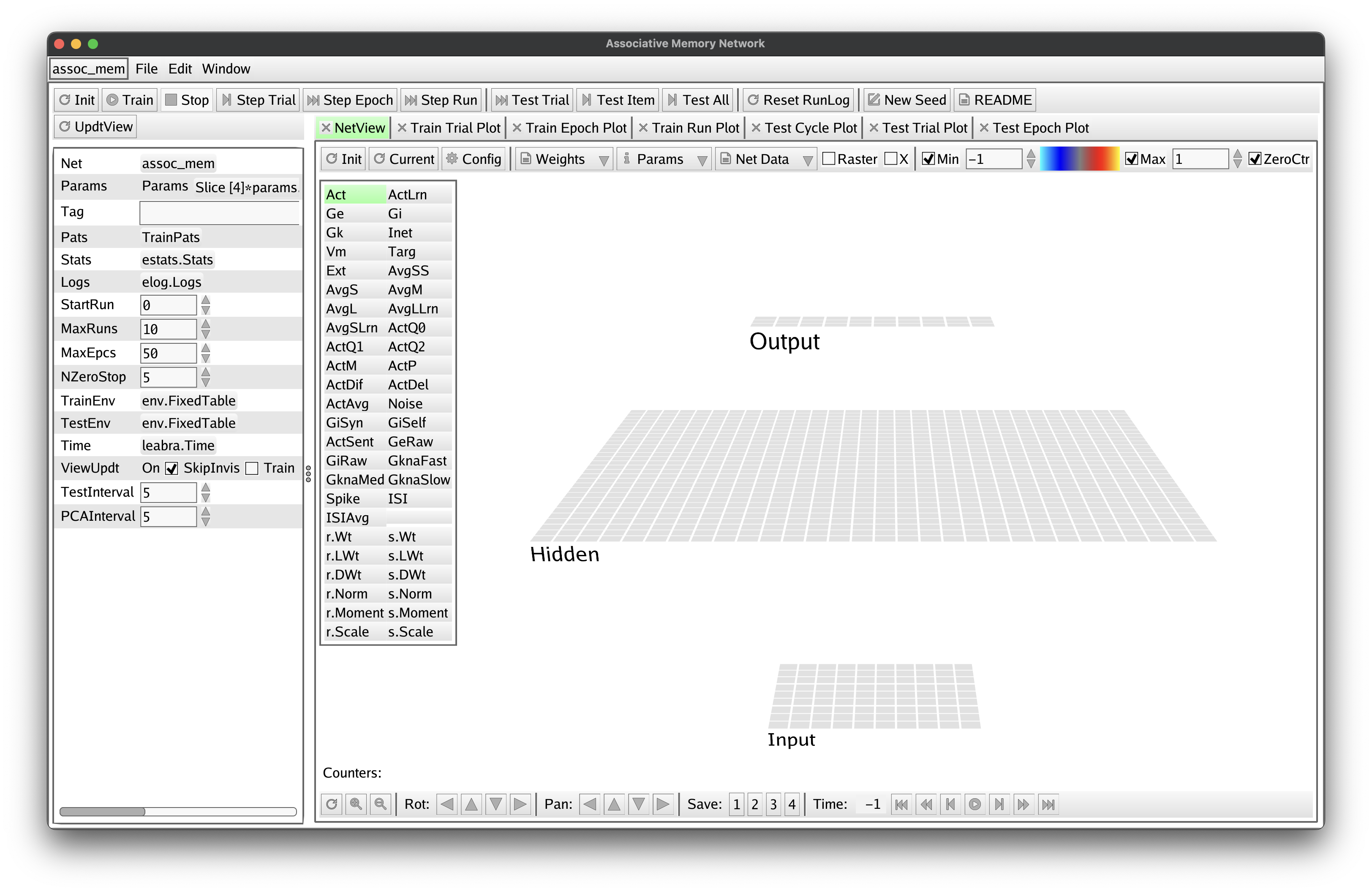}}{Associative memory network simulation.}{fig:2}

The final architecture or the associative memory model network consists of one input layer, a hidden layer, and an output layer. The implementation of this model takes inspiration from previous simulations developed as part of the \textit{Computational Cognitive Neuroscience} textbook by \citet{OReillyMunakataFrankEtAl12}. The model was designed to handle addition equations with operands ranging from 0–9.

The input layer (the ``Equation'') is a 4D layer of size $3\times1\times3\times10$. Both of the first two rows represent the first two operands in the equation. They consist of 30 units since that is the size of the embeddings produced by the numerical embedding network. The third row is simply a one-hot encoding of the operator: \texttt{0001} for addition, \texttt{0010} for subtraction, \texttt{0100} for multiplication, and \texttt{1000} for division.\footnote{Note that the current model is only trained on addition tables. As discussed in section 5, future work may focus on expanding the scope of the training data.} 

The hidden layer is a 2D layer of size $30\times30$. This is an arbitrarily large number; the magnitude of the size of this layer, rather than the exact number of units, is of most importance. As will be discussed later, the size of this layer influences the performance of the model. The magnitude of the size of this layer is crucial to allow the network to accommodate to the large training dataset.

The output layer is a 2D layer of size $3\times10$, which again corresponds to the size of the embeddings in use. This layer simply represents the output of the equation.

Together, these layers model a simple associative memory network. The architecture of this model influences its ability to generalize to equations it has not been taught yet, as well as its ability to sequentially learn new equations, as will be discussed in section 5.

\subsection{Training Procedure}

Using a series of Python notebooks run via Google Colab, I generated training datasets in the form of \texttt{.tsv} files for both the numerical embedding network and the associative memory model. The dataset for the numerical embedding network is simply a list of 82 rows one one-hot encodings for numbers 0–81. The dataset for the associative memory model uses the embeddings generated by the numerical embedding network to represent, in each row, the two operands and the output (the operator is simply a one-hot encoding, as previously mentioned).

Both networks were trained until the \texttt{PctErr} metric reached 0 for 5 epochs. For the numerical embedding network, this was reached at around 40 epochs across multiple trial runs (see \figref{3}). For the associative memory model, this was reached between 120–150 epoch across different trial runs (see \figref{4}). Because 0\% error was achieved in both models, we can conclude that they both successfully learned their entire training datasets.

\insertfig{\includegraphics[width=\textwidth]{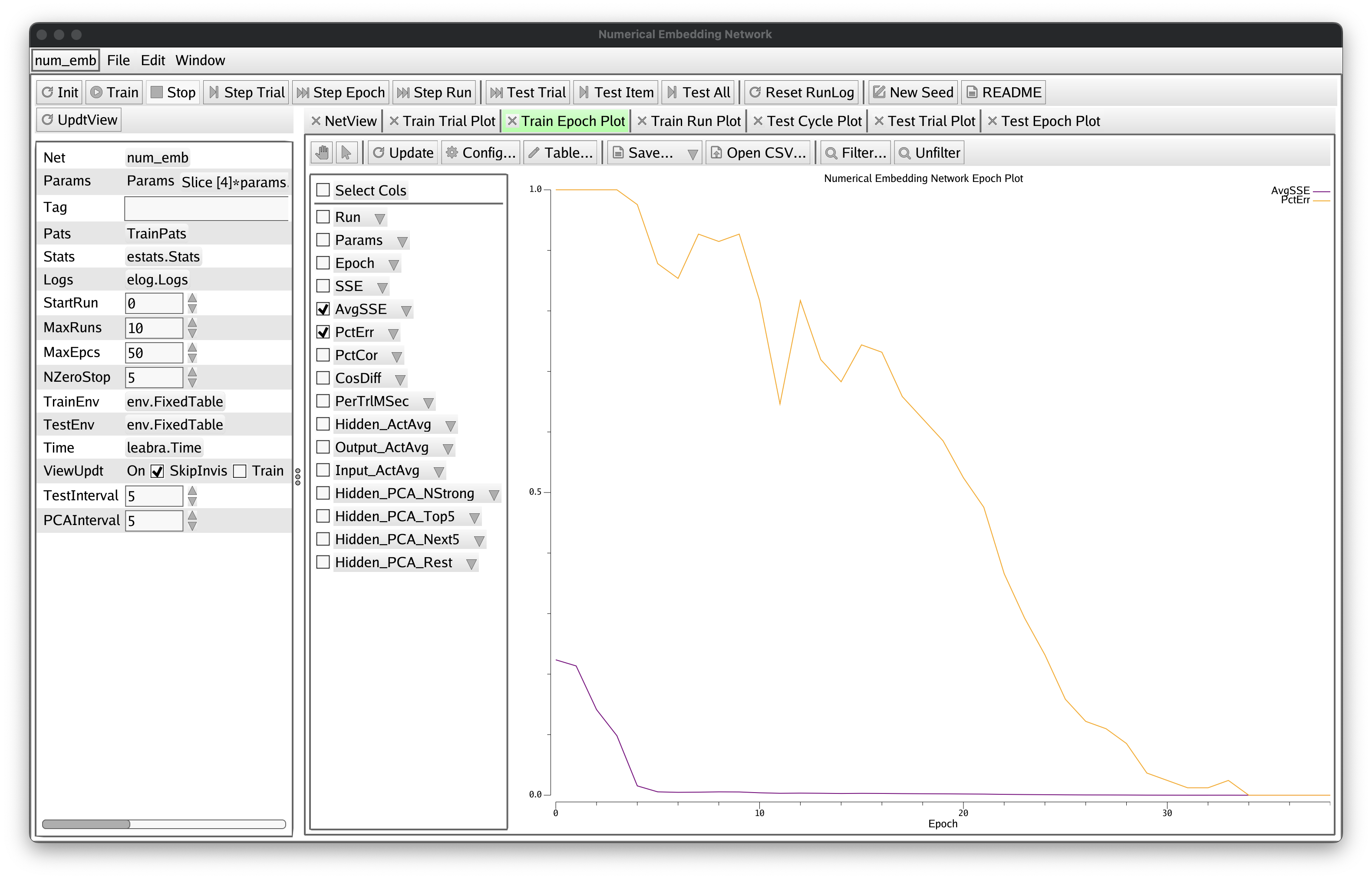}}{Train Epoch Plot for numerical embedding network.}{fig:3}
\insertfig{\includegraphics[width=\textwidth]{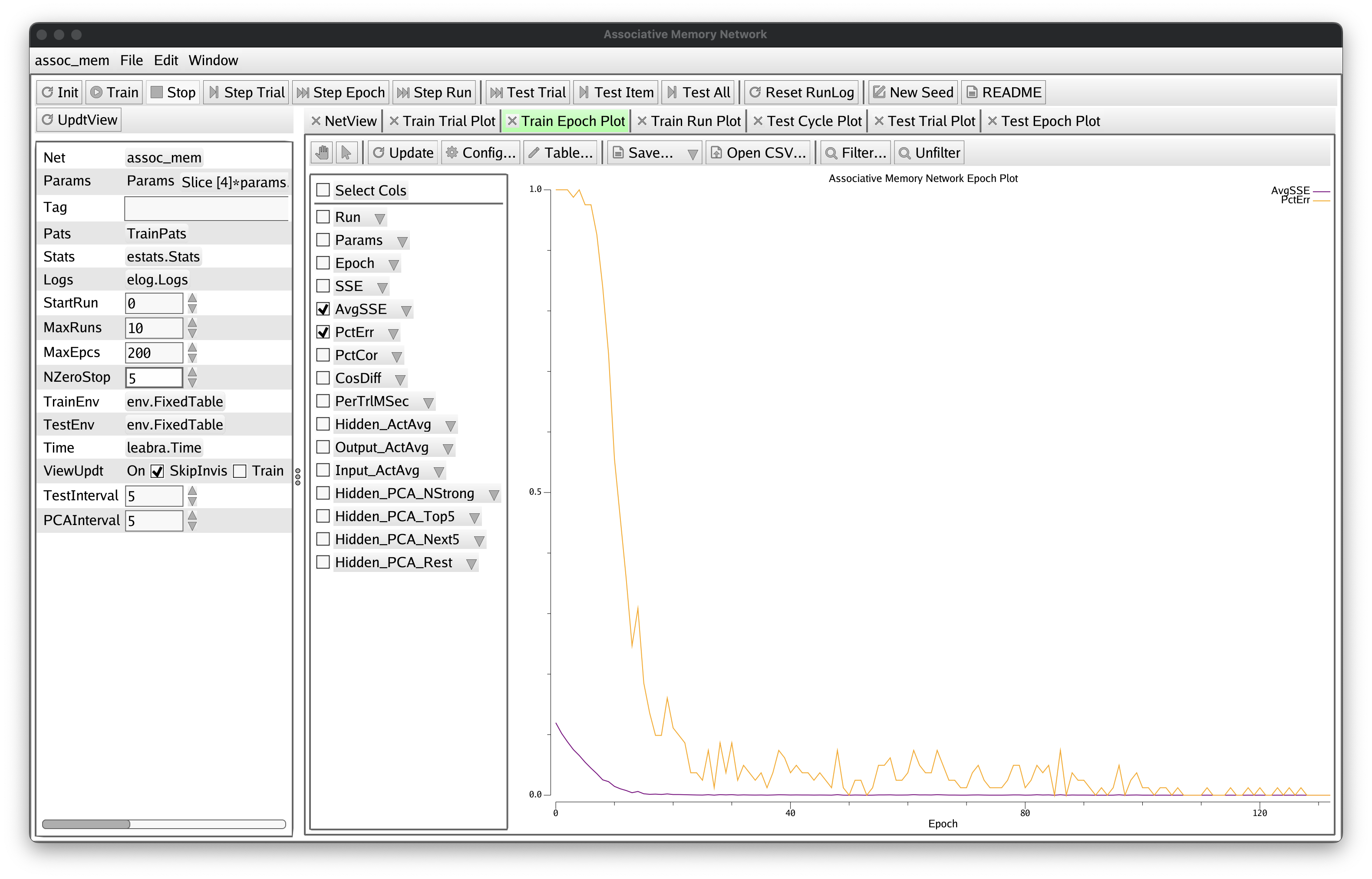}}{Train Epoch Plot for associative memory network.}{fig:4}

\subsection{Experimental Design}

\subsubsection{Simulating Dyscalculia via Neuronal Lesioning}

Dyscalculia is a neurodevelopmental disorder characterized by difficulty in understanding and processing numerical information, often resulting in impaired arithmetic abilities. \textit{Operational dyscalculia} in particular is ``manifested by impaired ability to perform mathematical operations, add, subtract, multiply, and divide'' \parencite{Viktorin2023}.

Taking inspiration from work performed by \citet{Hinton1991-cz} to simulate acquired dyslexia by lesioning neurons, I aim to simulate dyscalculia through lesioning the neural network models in this cognitive architecture. I experimentally lesioned neurons in both the numerical embedding network and the associative memory model to investigate two hypothetical causes of dyscalculia: failure to properly encode numbers into sparse distributed representations, and failure to develop associations between equation inputs and outputs. By inducing controlled lesions within these models, I seek to investigate how specific disruptions to the neural network affect its ability to learn and perform arithmetic tasks, thus providing insights into the underlying mechanisms of dyscalculia.

\subsubsection{Catastrophic Interference}

Catastrophic interference is a phenomenon observed in connectionist models where the learning of new information disrupts previously learned information, leading to a degradation in performance on tasks that were previously mastered. Catastrophic interference primarily arises due to the overwriting of previously learned information when new information is learned and interference between representations \parencite{McCloskey1989}. In the context of this study, catastrophic interference may occur when training the connectionist model sequentially on new addition and multiplication tables.

To investigate whether catastrophic interference would impact the associative memory model involved in this study, I conducted experiments where the connectionist model was trained on additional arithmetic tasks after mastering initial addition and multiplication tables. I tested the models on the initial dataset before and after training on additional data to investigate whether performance on these tasks diminished, thereby identifying the impact of catastrophic interference in this architecture.

\subsubsection{Generalization}

The associative memory model involved in this study was trained to learn equations from a provided dataset. While it is trivial to confirm or refute whether the model is able to successfully learn associations between equation inputs and outputs in the training data, if the model is able to generalize to tasks beyond what it has been trained on, that would suggest the model successfully ``learns'' to perform arithmetic beyond rote memorization of specific equations.

To investigate whether this model would be capable of generalization, I selectively excluded some equations from the dataset the associative memory model was trained on. I then tested the model on these excluded equations to determine whether it successfully solved them.

\section{Results}

\subsection{Simulating Dyscalculia via Neuron Lesioning}

Beginning at the entry point of the cognitive architecture, I experimented with lesioning the SDR embedding layer of the numerical embedding network to see if the model could still successfully learn to encode each number 0–81. If the model failed to do so, this would cause errors in training the associative memory model since different numbers would be confounded with each other, thereby serving as a cause of operational dyscalculia.

I ran each experiment for a maximum of 200 epochs, and recorded the final percent error achieved by each size, as shown in \tableref{1}. Intriguingly, the model still successfully learns the entire dataset even when 40\% of the neurons are lesioned, implying that the embedding layer of this network could be shrunk by around 40\% without losing the ability to produce SDRs for all of the training examples. However, any more reduction in dimensionality will lead to a reduction in performance, as at around 45\% lesion proportion, the model quickly begins to falter.

\begin{table}[H]
\centering
\begin{tabularx}{0.8\textwidth} { 
  >{\centering\arraybackslash}X 
  | >{\centering\arraybackslash}X 
  | >{\centering\arraybackslash}X }     
        \textbf{Lesion Proportion} & \textbf{Percent Correct} &  \textbf{\# of Epochs} \\ \hline
        0\% & 100\% & 38 \\
        40\% & 100\% & 74 \\ 
        45\% & 100\% & 200 \\ 
        50\% & 98.78\% & 200 \\ 
        60\% & 62.20\% & 200 \\ 
\end{tabularx}
\vspace{10pt}
\caption{Results of lesion experiments on the numerical embedding network.}
\label{tab:1}
\end{table}

The same experiment procedires were performed on the associative memory network, except each trial run was permitted to train for a maximum of 400 epochs (see \tableref{2}). Once again, the model quickly begins to fail to learn after more than 10\% lesioning of the hidden layer.

\begin{table}[H]
\centering
\begin{tabularx}{0.8\textwidth} { 
  >{\centering\arraybackslash}X 
  | >{\centering\arraybackslash}X 
  | >{\centering\arraybackslash}X }
        \textbf{Lesion Proportion} & \textbf{Percent Correct} &  \textbf{\# of Epochs} \\ \hline
        0\% & 100\% & 132 \\
        10\% & 100\% & 157 \\
        12\% & 98.77\% & 400 \\
        50\% & 51.85\% & 400 \\ 
\end{tabularx}
\vspace{10pt}
\caption{Results of lesion experiments on the associative memory network.}
\label{tab:2}
\end{table}

The rate at which the performance of these models begins to degrade once lesioned to a certain size demonstrates the precision required by these networks, and showcases how even relatively minimal neuronal damage can result in difficulties with arithmetic.

\subsection{Catastrophic Interference}

To test catastrophic interference within this cognitive architecture, I first trained the model on an addition table for the number 1 (i.e. $1 + 1$, $1+2$, etc.). I then tried training the model on an addition table for the number 2, thereby replicating sequential learning. To do so, I split the complete training dataset into individual training sets for each table.

After training the associative memory model on the initial training set, it reached a 0\% error rate when tested on these examples. However, when tested on the original examples after being trained on the new training data, the model exhibited an error rate of 100\% (see \figref{5}). This means that training the model on the new training data nearly entirely overwrote the model's memory of previous training data.

\insertfig{\includegraphics[width=\textwidth]{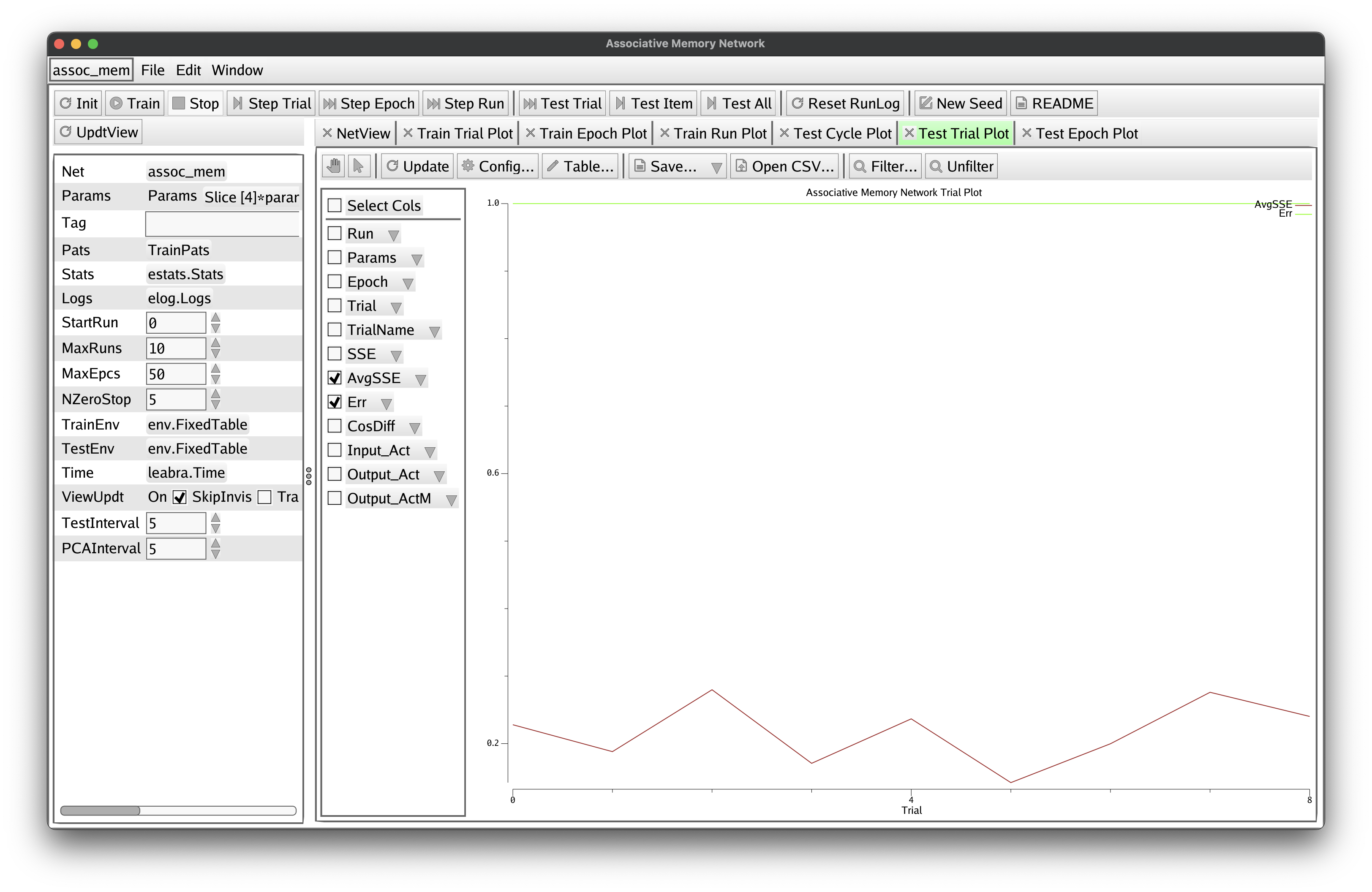}}{Testing results on addition table for 1 following training on 2.}{fig:5}

As a result of this experiment, we can conclusively state that this cognitive architecture is heavily impacted by catastrophic interference. This suggests that the associative memory model is not entirely robust and is susceptible to the disruption or erasure of previously learned information.

\subsection{Generalization}

To test the model's generalization capabilities, I randomly split up the training data into training and testing datasets. 20\% of the entire training data was marked as testing examples, and the remaining 80\% were marked as training examples.

After successfully training the model on exclusively the training examples, I tested all of the testing examples to see whether the model could generalize to these examples. Unsurprisingly, the model failed all testing examples and only succeeded on training examples (see \figref{6}). Because the model fails to generalize to these examples, this suggests there is little to no overlap between the hidden representations of different equations in this current model.

\insertfig{\includegraphics[width=\textwidth]{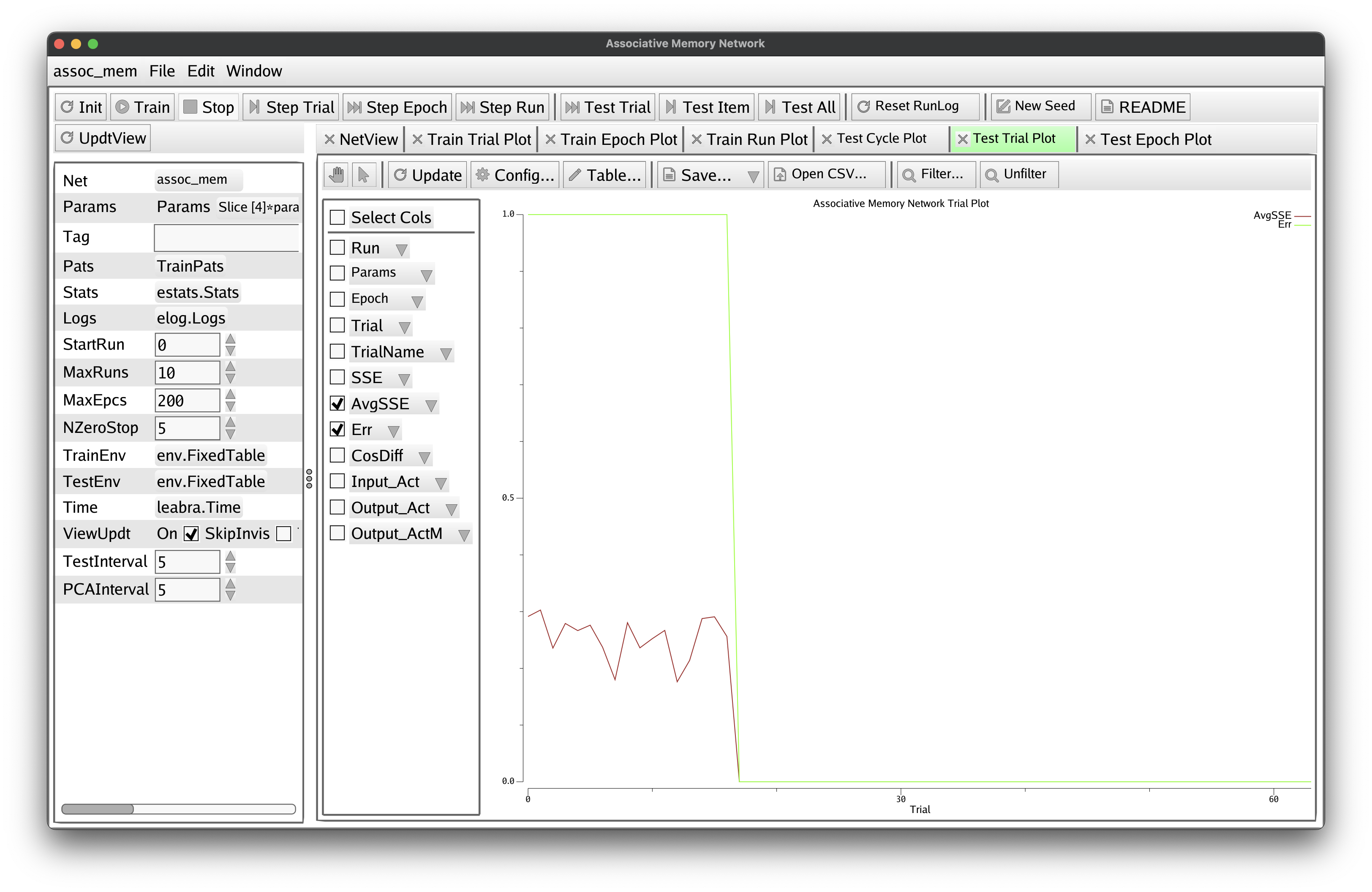}}{Testing results on testing then training data.}{fig:6}
    
\section{Discussion}

\subsection{Interpretations of Results}

The results of the experiments shed light on the capabilities and limitations of the proposed cognitive architecture for learning arithmetic operations. The successful training of both the numerical embedding network and the associative memory model demonstrates the feasibility of simulating arithmetic learning within a connectionist framework. However, several key findings warrant further discussion.

The experiments simulating dyscalculia through neuronal lesioning empirically demonstrate the precision and intricacy involved in these neural networks—they must simultaneously strike a balance between efficient neural representations and upscaling to accomodate learning new examples. There was a critical threshold beyond which the models' performance deteriorated rapidly, indicating the importance of preserving the integrity of neural representations for effective arithmetic learning.

The inability of the associative memory model to generalize to unseen data underscores the limitations of the current architecture in capturing abstract arithmetic concepts. The lack of overlap between hidden representations of different equations suggests that the model's learning is highly specific to the training data, hindering its ability to generalize to novel tasks. This limitation is consistent with findings in cognitive neuroscience, where studies have emphasized the importance of abstract representations in mathematical cognition.

For instance, Stanislas Dehaene, a prominent cognitive neuroscientist, has extensively studied the neural mechanisms underlying numerical cognition and arithmetic processing. Dehaene's work emphasizes the role of abstract numerical representations, such as number symbols and magnitude codes, in mathematical reasoning \parencite{Dehaene1999}. According to Dehaene's triple-code model, numerical information is processed through three distinct systems in the brain: a verbal number system, a visual number form area, and a magnitude comparison network, mechanisms which research suggest may take place in the intraparietal sulcus \parencite{Dehaene2003}. These systems interact to facilitate various arithmetic operations and numerical tasks beyond simply rote memorization.

The lack of abstract representations also may contribute to the model's susceptibility to catastrophic interference. Catastrophic interference is almost exclusively present in ANNs—the neuroplasticity of biological neural networks present in the human brain makes it much less susceptible to this shortcoming. This neuroplasticity enables the formation of abstract representations that capture the underlying principles and relationships within a domain, facilitating generalization and transfer of learning. In contrast, ANNs typically lack the same degree of flexibility and adaptability, making them more prone to catastrophic interference when exposed to new training data,

\subsection{Future Directions}

While the current study provides valuable insights into the cognitive mechanisms underlying arithmetic learning, several limitations merit consideration. Firstly, this study only investigates training this architecture on addition tables. However, the scope of this training data could be trivially expanded in future work. Doing so would involve creating new training examples for the associative memory model to learn from, which may require expanding the size of the hidden layer to accommodate for the growth in the size of the training dataset.

Moreover, the absence of end-to-end experiments integrating changes in the numerical embedding model with the associative memory network limits the comprehensive understanding of the architecture’s performance. Future research could explore the interplay between different components of the cognitive architecture to elucidate their collective contribution to learning and memory processes.

Additionally, the reliance on simplified simulations may oversimplify the complexity of real-world arithmetic learning. As previously mentioned, arithmetic learning likely involves more than associating equation inputs with outputs since no mathematical concepts are developed through this process \parencite{Dehaene1999}. Future work could extend the scope of this cognitive architecture to encompass a broader range of concepts in mathematical cognition, such as magnitude comparison, to improve the model's number sense and potentially increase its ability to generalize beyond its training examples.

In conclusion, this current study represents a step toward understanding arithmetic learning within a connectionist framework. Continued interdisciplinary research should look to refine and extend the proposed cognitive architecture for broader applications in cognitive science and artificial intelligence.

\pagebreak
\printbibliography

\end{document}